# Analysis of the manufacturing variation in a coupled micro resonators array based on its designed values and measured eigenfrequencies

Xueyong Wei, Nor Hayati Saad and Mike CL Ward

School of Mechanical Engineering, University of Birmingham, B15 2TT, UK

Tel: 44-121-4144226, E-mail address: x.wei.2@bham.ac.uk

**Abstract:** Manufacturing variation in the micro fabrication process inevitably alters the size of the designed structure and thereafter influences the performance of devices to some extent. A good knowledge of this effect will help the design and optimization of the devices. In this paper, we propose an analytical model to investigate the effect of manufacturing variation on the individual resonator in a chain of coupled micro mechanical resonators. It is also used to estimate the variation caused by the manufacturing process based on the designed values and the measured eigenfrequencies.

## 1. Introduction

Micromechanical resonators have been widely studied in various fields due to their small size, low power consumption and high sensitivity [1-3]. Usually, a resonator of single degree-of-freedom (DOF) measures a unique physical change (e.g., pressure, mass or stress) by tracking the shifting of its natural frequency. In recent few years, micro resonators of the multiple DOF architecture have been proposed to detect multiple physical changes in parallel [4-5].

However, all the devices need to be characterized before the practical applications. One of the reasons is that the inevitable manufacturing variation taken place in the micro fabrication process will result in a discrepancy between the measured performance and the design expectation [6]. The previous numerical study showed that a small amount of manufacturing variation still have an impact on the performance of coupled micro resonators array, through affecting the stability of mode shape and the measurability of the frequency response [7]. In order to use such an array of coupled micro resonators for sensing applications, one needs to characterize the fabricated micro devices by taking into account the manufacturing variation. Furthermore, a good knowledge of the possible range of manufacturing variation will help design a robust device [8].

In the following paragraphs, we propose a mathematical model to analyze the manufacturing variation in a chain of five weakly coupled micro mechanical resonators based on their designed values and the measured frequency responses. First, an indicator characterizing the effect of manufacturing variation on the performance of resonators is defined using a single DOF micro resonator as an example. Then, the mathematical model is introduced and the eigenvalues of the fabricated system is analyzed using perturbation theory. In this model, the eigenvalue of the perturbed system is linked with the designed value through a systematic matrix and the manufacturing variations. Finally, the experimental results are used to estimate the manufacturing variation, followed by some discussions.

## 2. Analytical Model

In principle, a single DOF micro mechanical resonator can be modelled using a mass-spring system including a dashpot and its equation of motion can be simplified in (1).

$$\ddot{x} + \frac{1}{Q}\dot{x} + \omega_0^2 x = F \cos \Omega t \qquad (1)$$

Where, $\omega_0 = \sqrt{\frac{k}{m}}$, $\Omega$, and $Q = \frac{m\omega_0}{c}$ are respectively the natural frequency of the system and the driving frequency of the external excitation force.

The resonant frequency of such a system is given by (2).

$$\omega_r = \omega_0 \sqrt{1 - \frac{1}{2Q^2}} \qquad (2)$$

Usually, Q factor of micro or nano mechanical resonators tested in vacuum is several thousand or higher. Therefore, (2) indicates that the natural frequency of a resonator can be characterized by using the resonant frequency identified from its frequency response curve, but the measured natural frequency is usually different from the designed value due to the processing variations. For example, a typical frequency response of a single DOF micro mechanical resonator compared with its designed curve is shown in figure 1.

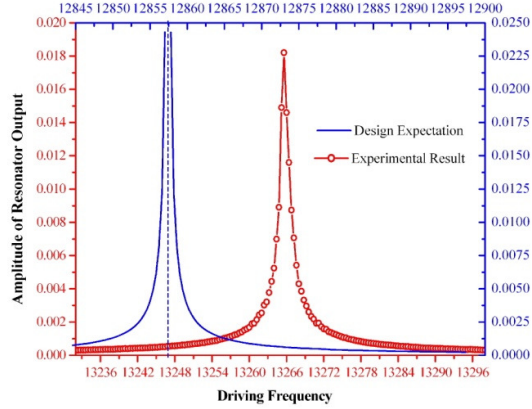

Figure 1 A typical frequency response of a single micro resonator compared with its designed curve

The change of natural frequency is mainly originated from the variation of the spring stiffness or the mass of the resonator and the relationship between them is governed by (3).

$$\frac{\Delta \omega_0}{\omega_0} = \frac{1}{2}\left(\frac{\Delta k}{k} - \frac{\Delta m}{m}\right) \quad (3)$$

Apparently, the relative change of the natural frequency is a comprehensive index charactering the variation of mass and spring stiffness. To make the analysis of coupled resonators easier, we define the index in a different form as below.

$$\left(\frac{\omega_r}{\omega_d}\right)^2 = 1 + \alpha_V \quad (4)$$

Where, $\omega_r$, $\omega_d$ are respectively resonant frequency and designed natural frequency of each element in the coupled resonators and $\alpha_V$ is the characterization index of variation. For example, a micro resonator tested in a vacuum (0.1Pa) shows the resonant peak at 13265.6Hz with a Q factor of 11377, while the designed natural frequency is 12857.4Hz as shown in figure 1. Accordingly, its index characterizing the manufacturing variation is that $\alpha_V = (\frac{13265.6}{12857.4})^2 - 1 = 6.45\%$.

In fact, all physical parameters of the coupled microresonator could be affected by the manufacturing process. Therefore, for a chain of N coupled identical resonators as designed in figure 2, the equations of motions are given as in (5).

$$m_j\ddot{x}_j + c_j\dot{x}_j + k_j x_j = k_{cj}(x_{j-1}-x_j) + k_{cj+1}(x_{j+1}-x_j), \quad 1 \le j \le N \quad (5)$$

Where, $m_j$, $c_j$, $k_j$ and $k_{cj}$ are the mass, damping, onsite spring stiffness and coupling spring stiffness of the fabricated micro resonator respectively and they are usually not the same as the designed values. The boundary conditions are $x_0 = x_{N+1} = 0$

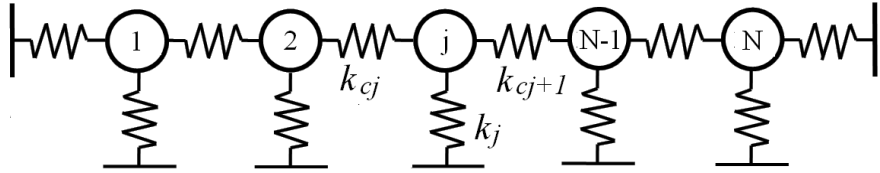

Figure 2 Schematic of a chain of N coupled micro resonators

Considering all the resonators are of same geometric design, it is naturally assumed that all damping coefficiencies are the same and that their effect on the frequency is neglected in the following analysis according to (2). Using the nominal design values (mass $m_0$, onsite spring stiffness $k_0$ and coupling spring stiffness $k_c$), transformations $\omega_j = \sqrt{\frac{k_j}{m_j}}$ and $\tau = \omega_0 t$, (5) can be rewritten as below.

$$\ddot{x}_j + x_j\left(\frac{\omega_j}{\omega_0}\right)^2 = \left(\frac{\omega_j}{\omega_0}\right)^2 \left(\frac{k_{cj}}{k_j}(x_{j-1}-x_j) + \frac{k_{cj+1}}{k_j}(x_{j+1}-x_j)\right), \quad 1 \le j \le N \quad (6)$$

If we assume that the manufacturing variation has the same effect on the stiffness of coupling spring and onsite spring, then let the stiffness variations of small order $\varepsilon$ as shown in Eq. (7).

$$\frac{k_{cj}}{k_c} = 1 + o(\varepsilon), \text{ and } \frac{k_i}{k_0} = 1 + o(\varepsilon) \quad (7)$$

Hence, the stiffness ratio between the coupling spring and onsite spring can be approximated as below.

$$\frac{k_{cj}}{k_j} = \frac{k_c}{k_0}\frac{k_{cj}}{k_c}\frac{k_0}{k_j} \approx \frac{k_c}{k_0}(1 - o(\varepsilon^2)) \quad (8)$$

This indicates that the manufacturing variation has little effect on the stiffness ratio in our weakly coupled system. Denote $\frac{k_c}{k_0} = \beta$, and use the definition in (4), then (6) is further simplified.

$$\ddot{x}_j + x_j(1 + \alpha_{v1}) = \beta(1 + \alpha_{v1})(x_{j-1} - 2x_j + x_{j+1}), \quad 1 \le j \le N \quad (9)$$

For simplicity, the collective motion of coupled resonators takes the form of $x_j = H_j e^{i\sqrt{\lambda}\tau}$, where $\lambda$ is the eigenvalue of the system. Substitute the definition into (9) and rearrange the equations into matrix form, we have

$$[M][H] = \lambda[H] \quad (10)$$

Where, M and H are respectively the eigenmatrix and eigenvector of the system described in (9).

In order to predict the frequency response of the system governed by (9), consider $\alpha_V$ is a small parameter, and replace $\alpha_{vj}$ with $\varepsilon\alpha_j$, then (10) can be written as follows for applying the perturbation method to calculate the system eigenvalues.

$$([M_0] + \varepsilon[M_{MV}])[H] = \lambda[H] \quad (11)$$

Where, $[M_0] = \begin{bmatrix} 1+2\beta & -\beta & 0 & 0 & 0 \\ -\beta & 1+2\beta & -\beta & 0 & 0 \\ \ldots & \ldots & \ldots & \ldots & \ldots \\ 0 & 0 & -\beta & 1+2\beta & -\beta \\ 0 & 0 & 0 & -\beta & 1+2\beta \end{bmatrix}$ represents the ideal case of N coupled resonators without any manufacturing variations and

$[M_{mv}] = \begin{bmatrix} \alpha_1(1+2\beta) & -\beta\alpha_1 & 0 & 0 & 0 \\ -\beta\alpha_2 & \alpha_2(1+2\beta) & -\beta\alpha_2 & 0 & 0 \\ \ldots & \ldots & \ldots & \ldots & \ldots \\ 0 & 0 & -\beta\alpha_4 & \alpha_4(1+2\beta) & -\beta\alpha_4 \\ 0 & 0 & 0 & -\beta\alpha_5 & \alpha_5(1+2\beta) \end{bmatrix}$ represents the effect of manufacturing variation.

For the ideal case $[M_0][H] = \lambda_0[H]$, the $p^{th}$ eigenvalue and its corresponding eigenvector of the equation can be solved analytically and given as follows.

$$\lambda_0^P = 1 + 4\beta \sin^2\left(\frac{p\pi}{2(N+1)}\right), 1 \leq p \leq N \tag{12a}$$

$$[H_0^p] = \left[\sin\left(\frac{p\pi}{N+1}\right), \sin\left(\frac{2p\pi}{N+1}\right), \sin\left(\frac{3p\pi}{N+1}\right), \cdots, \sin\left(\frac{Np\pi}{N+1}\right)\right], 1 \leq p \leq N \tag{12b}$$

Applying the perturbation method [9], the eigenvalues of (11) can be approximated as follows.

$$\lambda^P = \lambda_0^P + \varepsilon\lambda_{MV}^P = \lambda_0^P + \varepsilon[H_0^p][M_{mv}][H_0^p]^T / \|H_0^p\|^2 \tag{13}$$

It is clearly indicated in (13) that the eigenfrequencies of the coupled system can be predicted once the manufacturing variation is known. Similarly, if the frequency responses of the system are known, then (13) can be used to study the manufacturing variation as shown below.

$$\lambda - \lambda_0 = \varepsilon\lambda_{mv} = [D][MV]\begin{bmatrix}\varepsilon\alpha_1\\\varepsilon\alpha_2\\\cdots\\\varepsilon\alpha_{N-1}\\\varepsilon\alpha_N\end{bmatrix} = [D][MV]\begin{bmatrix}\alpha_{v1}\\\alpha_{v2}\\\cdots\\\alpha_{vN-1}\\\alpha_{vN}\end{bmatrix} \tag{14}$$

Where, 
$$\begin{cases}MV_{p,1} = (1+2\beta)\sin^2\left(\frac{p\pi}{N+1}\right) - \beta\sin\frac{p\pi}{N+1}\sin\frac{2p\pi}{N+1},\\ MV_{p,q} = (1+2\beta)\sin^2\left(\frac{pq\pi}{N+1}\right) - 2\beta\sin^2\left(\frac{pq\pi}{N+1}\right)\cos\frac{p\pi}{N+1}, 1 \leq p \leq N, 2 \leq q \leq N-1 \text{ and}\\ MV_{p,N} = (1+2\beta)\sin^2\left(\frac{Np\pi}{N+1}\right) - \beta\sin\frac{Np\pi}{N+1}\sin\frac{(N-1)p\pi}{N+1}.\end{cases}$$

the matrix [D] is a diagonal matrix with its element $D_p^{-1} = \frac{1}{\sum_{q=1}^N \sin^2\left(\frac{pq\pi}{N+1}\right)}, 1 \leq p, q \leq N$.

It is indicated in (14) that the manufacturing variation can be uniquely determined if [MV] is a non-singular matrix. However, the rank of [MV] is (N+1)/2 when N is an odd number or N/2 when N is an even number, as the values in the $p^{th}$ column in [MV] equal to that in the (N+1-$p$)$^{th}$ column, which means that the manufacturing variation of the $p^{th}$ resonator has the same effect as that of the (N+1-$p$)$^{th}$ resonator does on the eigenvalues of the system. This also can be explained simply as the effect of symmetry of the chain about its middle resonator. Based on the designed values $\lambda_0$ and the measured eigenvalues $\lambda$, however, the sum of manufacturing variation pairs $(\alpha_{vp}, \alpha_{vN+1-p})$ can be figured out using the multiple regression analysis. In the next section, an example of applying above theory into five weakly coupled micro resonators is given based on experimental results.

## 3. Experiment and Discussion

In our studies, single and multiple DOF micro mechanical resonators were fabricated using Silicon on Insulator (SOI) technology that has been widely used in the research. Figure 3 shows the SEM image of a chain of coupled micro resonators, where the individual resonator highlighted by the dotted line is coupled by the S shape beam (S2).

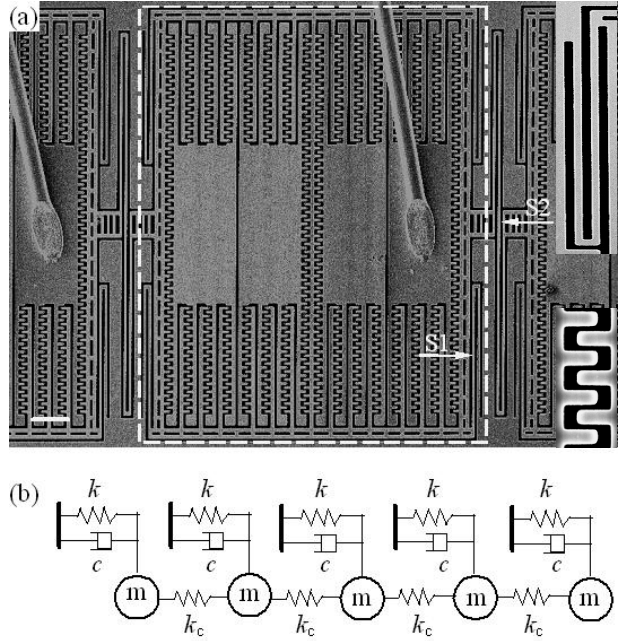

Figure 3 (a) SEM image of a chain of five coupled micro resonators, and the insets show the zoom in images of the anchor spring and comb fingers. (b) A schematic model represents the array. $k$, $k_c$ and m denote the anchor spring (S1) coupling spring (S2) and mass of the resonators.

To test the mathematical model introduced in the previous section, the chain of five weakly coupled resonators was measured by recording the response of one element while sweeping the frequency of the driving force at another element. The frequency responses of the designed coupled resonators and the fabricated are shown in figure 4, and the designed eigenfrequencies $f_0$, the designed eigenvalues $\lambda_0$, the measured eigenfrequencies $f_1$ and the calculated eigenvalues $\lambda_1$ are listed in Table 1.

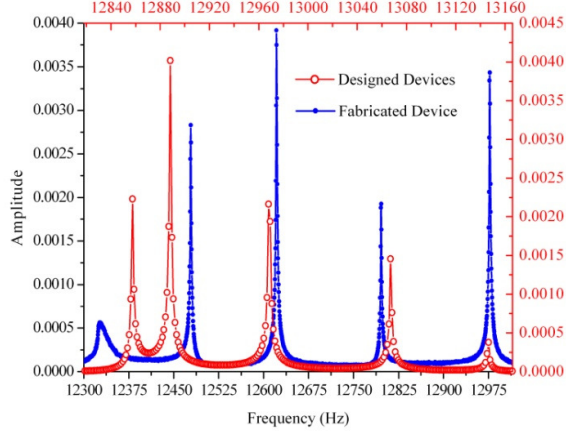

Figure 4 Frequency responses of the designed and microfabricated device as shown in Figure 3. The shift of the resonant peaks is due to the effect of the manufacturing variation on individual resonator.

Table 1 eigenfrequency and eigenvalue of the designed and microfabricated five weakly coupled micro mechanical resonators

|             | 1       | 2       | 3       | 4      | 5       |
|-------------|---------|---------|---------|--------|---------|
| $f_0$       | 12857.4 | 12888.3 | 12968   | 13067  | 13146.7 |
| $\lambda_0$ | 1.003   | 1.013   | 1.026   | 1.039  | 1.049   |
| $f_1$       | 12321.6 | 12478.8 | 12621.2 | 12796  | 12977.2 |
| $\lambda$   | 0.918   | 0.942   | 0.964   | 0.99   | 1.019   |
| $\lambda - \lambda_0$ | -0.085 | -0.071 | -0.062 | -0.049 | -0.03 |

Applying the theory detailed in Section 2 into this system, we obtain the following prediction of eigenvalue changes as shown in (15). The second term on the right-hand side of (15) is due to the fact that the resonators at both ends of the chain are not in connection with the substrate.

$$\lambda - \lambda_0 = \varepsilon \lambda_{mv} = \frac{1}{3}[\overline{MV}]\begin{bmatrix}\gamma_1\\\gamma_2\\\gamma_3\end{bmatrix} - \frac{2\beta}{3}\begin{bmatrix}\sin^2\left(\frac{\pi}{6}\right)\\ \ldots \\ \sin^2\left(\frac{q\pi}{6}\right) \\ \ldots \\ \sin^2\left(\frac{5\pi}{6}\right)\end{bmatrix} \quad (15)$$

Where, 
$$\begin{cases}\overline{MV}_{p,1} = (1+\beta)\sin^2\left(\frac{p\pi}{6}\right) - \beta \sin\frac{p\pi}{6}\sin\frac{2p\pi}{6}, \\ \overline{MV}_{p,2} = (1+2\beta)\sin^2\left(\frac{2p\pi}{6}\right) - 2\beta\sin^2\left(\frac{2p\pi}{6}\right)\cos\frac{p\pi}{6} \quad 1 \leq p \leq 5, \\ \overline{MV}_{p,3} = (1+2\beta)\sin^2\left(\frac{3p\pi}{6}\right) - 2\beta\sin^2\left(\frac{3p\pi}{6}\right)\cos\frac{p\pi}{6}\end{cases}$$

$\gamma_1 = \alpha_{v1} + \alpha_{v5}, \gamma_2 = \alpha_{v2} + \alpha_{v4}$, and $\gamma_3 = \alpha_{v3}$.

Treat the three columns of matrix $[\overline{MV}]$ as the values of the three independent variables and the $\gamma_1, \gamma_2$ and $\gamma_3$ as three coefficients of the variables, then Eq.(15) becomes the target function to be optimized to match the measured results of eigenvalue difference as listed in Table 1. Applying the multiple regression analysis to the problem, we obtained the results as follows.

$\gamma_1 = -0.103, \gamma_2 = -0.1053, \gamma_3 = -0.0553$

If we assume the manufacturing variation has the same effect on the resonators in symmetry about the middle one of the chain, then we have $\alpha_{v1} = -5.15\%, \alpha_{v2} = -5.26\%, \alpha_{v3} = -5.53\%, \alpha_{v4} = -5.26\%, \alpha_{v5} = -5.15\%$.

As a deduction of Eq. (3) and Eq. (4), we have

$$\frac{\Delta\omega_0}{\omega_0} \approx \frac{1}{2}\alpha_V, \text{ or } \alpha_V \approx \frac{\Delta k}{k} - \frac{\Delta m}{m} \qquad (16)$$

Therefore, the effective spring stiffness may decrease relatively 5% or the effective mass may increase relatively 5% when there is only one parameter change. In fact, both parameters are affected by the manufacturing process. As one can see from the insets in figure 3, the corner of the anchor springs and comb fingers is rounded. The microscopic measurement of the devices further confirms that the real dimensions are different from the designed values depending on their original dimensions and their shape. In the manufacturing process, there are many factors that can influence the designed structures. For example, a slightly over exposure will result in the distortion of small structures during the photolithography process and narrow gaps are etched faster than larger ones during the deep reactive ion etching (DRIE). In addition, metallization of the fabricated device for wire bonding will add an extra weight onto the micro resonators.

To verify the estimation of manufacturing variation given above and understand the possible causes of the variation, the dimensions of the fabricated devices were computed using the SEM figures and the image processing tool in Matlab. In the analyzed figure, each pixel represents a length of 296nm and the width and length of the structure are calculated by counting the pixels. Based on the average of the measured values, the effective stiffness of the resonator was simulated using Finite Element Analysis package (Comsol Multiphysics) and the mass was calculated by integrating the density over the area. As a result, the effective stiffness is 19.979 N/m for the designed resonator and 21.297 N/m for the fabricated one. The relative mass change can also be approximated by the area change as the thickness and the density of the devices are constant and it is about 13.129% as shown below.

$$\frac{\Delta m}{m} = \frac{(\Delta A_{finger} + \Delta A_{anchor} + \Delta A_{frame})}{A} \approx \frac{7.436\text{e}4 \text{ μm}^2 - 6.573\text{e}4 \text{ μm}^2}{6.573\text{e}4 \text{ μm}^2} = 13.13\%$$

Therefore, the index $\alpha_v$ for the fabricated device can be calculated according to Eq. (16) and it gives that $\alpha_v = \frac{21.297 - 19.979}{19.979} - 13.13\% = -6.53\%$, which is close to the theoretical prediction. It indicates that the model can be used to estimate the effect of manufacturing variation on the coupled resonators.

## 4. Conclusions

There are many inevitable factors in microfabrication process affecting the final dimensions of the products and thereafter their performance. In this paper, we propose a mathematical model to extract the manufacturing variation of a coupled micro resonator array. The results indicate that the effect of the manufacturing variation on each resonator in the coupled array is consistent at the same level. Combining the index for single and multiple DOF resonators, the results show that there is about ±3% variation of the designed frequency caused by the inevitable manufacturing variation. The model described in Section 2 also shows the possibility to estimate the frequency change once the variation is known.

## 5. Acknowledgement

This work is financially supported by the UK Engineering and Physical Sciences Research Council (EPSRC) and the authors are also grateful to the technical support from Dr. Steve Collins and Ross Turnbull for the electrical testing of the fabricated devices.